\documentclass[11pt]{article}
\usepackage[a4paper,margin=25mm]{geometry}
\usepackage{graphicx}
\usepackage{amsmath}
\usepackage{amssymb}
\usepackage{newtxtext}
\usepackage{newtxmath}
\usepackage[round,authoryear]{natbib}
\usepackage{authblk}
\usepackage{arydshln}
\usepackage[hidelinks]{hyperref}
\renewcommand{\vec}[1]{\boldsymbol{#1}}
\renewcommand{\d}{\textnormal{d}}
\newcommand{\D}{\textnormal{D}}
\numberwithin{equation}{section}
\title{Particle settling in turbidity currents: concentration-gradient-driven biased sampling}
\author[1]{Lianzheng Cui\thanks{Corresponding author: \href{mailto:lianzheng.cui22@imperial.ac.uk}{lianzheng.cui22@imperial.ac.uk}}}
\author[2,3]{Eric Climent}
\author[1]{Graham O. Hughes}
\author[1]{Maarten van Reeuwijk}
\affil[1]{Department of Civil and Environmental Engineering, Imperial College London, London SW7 2AZ, UK}
\affil[2]{Ayrton-Bl\'eriot Engineering Lab., IRL 2035 CNRS - Imperial College London, London SW7 2AZ, UK}
\affil[3]{Toulouse INP-ENSEEIHT, 2 rue Camichel - 31000 Toulouse, France}
\date{}
\begin{document}
\maketitle

\begin{abstract}
We investigate the mechanisms governing particle settling in turbidity currents using Eulerian--Lagrangian direct numerical simulations. The Eulerian carrier flow is driven either by solutal buoyancy or particle feedback, with the Lagrangian phase comprising passive tracers or inertial particles, respectively. The effective particle settling velocity is decomposed into a fluid velocity sampled at particle positions and a particle--fluid slip velocity. The Eulerian mean profiles of these velocities are obtained using a concentration-weighted average of the coarse-grained fields. The mean sampled fluid velocity is shown to be approximately equal to the ratio of the vertical turbulent flux of particles to their mean concentration and reflects biased sampling of upward turbulent fluctuations at particle positions, despite the zero Eulerian mean vertical fluid velocity. The passive-tracer cases show that the upward bias does not require particle inertia and arises from turbulent transport acting on concentration gradients, as it persists for inhomogeneous tracer seeding but disappears under uniform seeding. For the weakly inertial regime considered here, the upward bias dominates downward-directed biases associated with particle inertia. The mean slip velocity is well approximated by the terminal settling velocity predicted for a quiescent fluid. This is consistent with a leading-order balance between buoyancy and drag in the slope-normal direction. Modelling the sampled fluid velocity from the turbulent flux and using the slip-velocity approximation yield an Eulerian prediction for the settling velocity, in good agreement with the simulation data for the dilute, weakly inertial particles considered.
\end{abstract}

\section{Introduction}
\label{sec:Intro}

Turbidity currents are particle-laden gravity flows capable of sculpting submarine landscapes and posing hazards to seafloor infrastructure. These flows consist of a bottom inner layer analogous to an open-channel flow and a free-shear-like outer layer that grows through entrainment of ambient fluid \citep{luchi2018drivinglayer, wells2021ARFM}. Particle settling within these turbulent shear layers is a key process governing the evolution of the flow. However, even in the canonical problem of particles settling in homogeneous turbulence, the role of turbulence in modulating settling remains debated \citep[see, e.g.,][]{brandt2022ARFM}.

Most studies have reported enhanced settling for heavy (weakly) inertial particles in homogeneous isotropic turbulence \citep{maxey1987orginal, wang1993settling, aliseda2002settling, bec2014settling, rosa2016settling, tom2019settling}. This enhancement is commonly attributed to preferential sweeping, whereby particles are centrifuged out of vortices and oversample downwelling regions \citep{maxey1987orginal, wang1993settling}. \cite{bragg2021wallsweep, bragg2021wallsweep2} extended this mechanism to a turbulent boundary layer, demonstrating that preferential sweeping remains active throughout much of the depth. Near the bottom boundary, turbophoresis effects \citep{caporaloni1975turbo,reeks1983turbo} further enhance settling by driving particles from regions of high to low turbulence intensity.

In contrast to enhanced settling, some studies have suggested that turbulence may reduce particle settling. In homogeneous isotropic turbulence, the mechanism is often referred to as the ‘loitering’ effect \citep{nielsen1993loitering}. Although direct numerical simulation (DNS) links this reduced settling to nonlinear drag effects \citep{good2014settling}, its generality remains debated. In turbulent boundary layers, oversampling by particles of ejection (Q2) regions, characterised by upward motions of low-speed fluid, has been observed to counteract turbophoresis \citep{gao2024costalift}, yet the underlying dynamics also remain unclear.
 
In turbidity currents, particles can remain suspended by turbulence and drive the currents over long distances, which is referred to as auto-suspension \citep{bagnold1962auto,wells2021ARFM}. A distinguishing feature of these flows is the strong vertical variation of particle concentration, which decreases with height above the slope \citep[see e.g.][]{wells2021ARFM}. Statistically, such concentration gradients induce an upward turbulent flux under the gradient--diffusion hypothesis that counteracts settling \citep{rouse1939analysis}. However, it remains unclear how this process interacts with inertia-related settling-enhancement processes such as preferential sweeping and turbophoresis.

\begin{figure}
  \centerline{\includegraphics[scale =0.8]{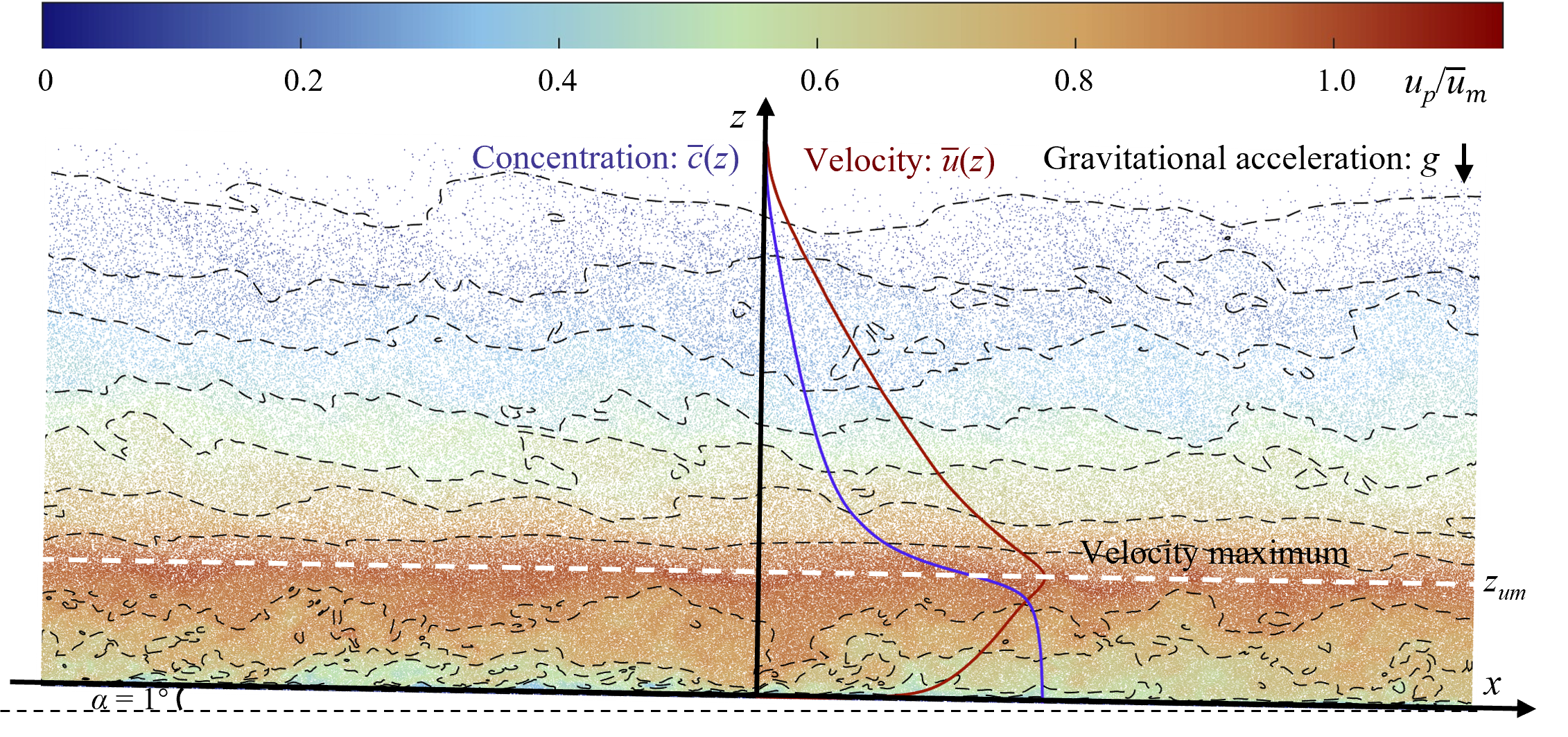}}
  \caption{Initial distribution of instantaneous streamwise particle velocity $u_p$ (indicated by the colour of each particle), normalised by the maximum mean streamwise fluid velocity $\overline{u}_m$. The profiles of the mean streamwise fluid velocity $\overline{u}$ (red solid line) and particle concentration $\overline{c}$ (blue solid line) are provided for reference. The white dashed line denotes the level of $z_{um}$, where $\overline{u}$ attains its maximum. }
  \label{fig:system}
\end{figure}

The objective of this study is to clarify the mechanisms that reduce particle settling velocities in turbidity currents. To this end, we perform two-way coupled Eulerian--Lagrangian DNS of mono-disperse turbidity currents with up to twenty-two billion particles, considering four particle sizes and a fixed initial concentration field. To distinguish the roles of particle inertia and concentration gradients, we also perform two solutally-driven simulations with passive tracers seeded either to match the heterogeneous buoyancy field or uniformly throughout the domain. We show that turbulent transport acting on concentration gradients generates an upward bias in the fluid velocity sampled at particle positions. This bias persists in the passive-tracer simulation when a concentration gradient is present and dominates classical inertia-related settling-enhancement mechanisms in the weakly inertial particle simulations. Ultimately, we aim to develop a theoretical framework that connects Eulerian turbulent transport with Lagrangian settling dynamics in turbidity currents which can be extended to a broader class of particle-laden flows.

\section{Methodology}\label{sec:method}
\subsection{Eulerian--Lagrangian governing equations}
\label{sec:equations}
In this study, we consider an Eulerian carrier-fluid phase coupled to two types of dispersed Lagrangian phases: passive tracers and inertial particles.  Before introducing either dispersed phase, we first perform DNS of a single-phase Boussinesq gravity current consisting of a negatively buoyant layer descending a slope of angle $\alpha=1^\circ$. The governing equations in the coordinate system shown in figure \ref{fig:system} are
\begin{align}
\label{eq:solutal continuity}
  \nabla\cdot{\vec u}  &=0,\\
  \label{eq:solutal fluid momentum}
{\partial_t \vec{u}}+{\vec u} \cdot \nabla {\vec u}&=-\nabla{p}^*+\nu\nabla^2\vec{u}+b\vec{e},\\
\label{eq:buoyancy transport}
{\partial_t b}+{\bf u} \cdot \nabla b &=\kappa\nabla^2 b.
\end{align}
Here, $\vec{u}=(u,v,w)$ is the fluid velocity and $b=(\rho_f-\rho)g/\rho_f$ is the buoyancy, where $\rho$ and $\rho_f$ denote the local and ambient fluid densities, respectively, and $g$ is the gravitational acceleration. The unit vector $\vec{e}=(-\sin{\alpha},0,\cos{\alpha})$, while $\nu$ and $\kappa$ are the kinematic viscosity and scalar diffusivity, respectively. The Schmidt number is fixed at $Sc=\nu/\kappa=1$. The kinematic pressure is ${p}^*=p/\rho_f-\vec g\cdot\vec x$, where $\vec g=-g\vec e$.
After the gravity current reaches a fully developed state, the resulting flow field is used to initialise the Eulerian--Lagrangian simulations. Figure \ref{fig:system} shows the resulting velocity field together with the profiles of Eulerian mean streamwise velocity $\overline{u}$ and particle concentration $\overline{c}$. The averaging operator $\overline{(\cdot)}$ is defined in \eqref{eq:Reave}. 

In the passive-tracer simulations, the solutally-driven current is supplemented with zero-inertia Lagrangian tracers that follow the local fluid velocity exactly:
\begin{equation}
    \frac{\d \vec{x}_p}{\d t} = \vec{u}(\vec{x}_p,t),
\end{equation}
where ${\vec x}_p=(x_p,y_p,z_p)$ denotes the position of a Lagrangian dispersed-phase element. Since the tracers are passive and exert no feedback on the fluid, the governing equations for the fluid phase remain those given by (\ref{eq:solutal continuity}--\ref{eq:buoyancy transport}). 

For the inertial-particle simulations, the solutal concentration field is replaced by a suspension of mono-disperse particles. We consider relatively small and dilute particles as discussed in \S\ \ref{sec:setup}. The motion of the particles is modelled following \cite{maxey1983equation, magnaudet2000forces}:

\begin{align}
\label{eq:dxdt}
\frac{\d \vec{x}_p}{\d t} &=\vec{v}(\vec{x}_p,t),\\
    m_p\frac{\d \vec{v}}{\d t}&=\vec{ F}_d + {\vec F}_{l}+ {\vec F}_{ad} + {\vec F}_{pv} + (m_p- m_f) \vec{g},
   \label{eq:dvdt}
\end{align} 
where $m_p$ and $m_f$ are the mass of the particle and the displaced fluid, respectively, and $\vec{v}=(u_p,v_p,w_p)$ denotes the particle velocity. The force $\vec{ F}_d$ is the drag, ${\vec F}_{l}$ is the shear-induced lift modelled using the Saffman formulation \citep{saffman1965lift}, ${\vec F}_{ad}$ is the added mass force, ${\vec F}_{pv}$ is the force that would apply on a fluid element at the particle position \citep{maxey1983equation}, where
\refstepcounter{equation}\label{eq:forces}
\begin{equation*}
\begin{aligned}
{\vec F}_d &= \frac{m_p(\vec{u}-\vec{v})}{\tau_{st}}\mathcal{F}(Re_p),
&\quad
{\vec F}_{l} &= m_f\frac{3C_{\textnormal{saff}}}{2\pi d_p}|\vec{u}-\vec{v}|
\sqrt{{\nu}|\vec{\omega}|}
\frac{(\vec{u}-\vec{v})\times \vec{\omega}}
{|(\vec{u}-\vec{v})\times \vec{\omega}|},
\\
{\vec F}_{ad} &= m_fC_M\left(\frac{\D \vec{u}}{\D t}-\frac{\d \vec{v}}{\d t}\right),
&\quad
{\vec F}_{pv} &= m_f\frac{\D \vec{u}}{\D t}.
\end{aligned}
\tag{\theequation a-d}
\end{equation*}
Here, $Re_p = {d_p|\vec{u}-\vec{v}|}/{\nu}$ is the particle Reynolds number, $\mathcal{F}(Re_p)=(1+0.15Re_p^{0.687})$ is the nonlinear drag correction modelled using the correlation of \cite{schiller1933Recor}, $\vec{\omega}=\nabla\times\vec{u}$ is the fluid vorticity, $C_{\textnormal{saff}} = 6.46$ is the Saffman constant, and $C_M=1/2$ is the added-mass coefficient for spherical particles. 

The Basset history force is omitted from \eqref{eq:dvdt}. Its magnitude relative to the quasi-steady drag is generally expected to scale with the particle diameter $d_p$ relative to the Kolmogorov length scale \citep[see, e.g.,][for a proportionality coefficient of 0.35]{daitche2015history_dp_eta}; see also \citet{ling2013history_dp_eta,gode2023history_dp_eta}. This ratio remains small for the cases considered here (see table \ref{tab:sim}). Although some history-induced changes in particle dynamics, such as preferential concentration, may also depend on the density ratio $\rho_p/\rho_f$ \citep{daitche2015history_dp_eta}, these effects are secondary in the weakly inertial regime considered here, as discussed in \S\ref{sec:theory}. Moreover, the history force generally makes inertial particles follow the flow more closely \citep{daitche2011memory,daitche2014memory,daitche2015history_dp_eta}, while the present inertial-particle cases already resemble the passive-tracer cases (see \S\ref{sec:decomp}). Neglecting the history force is thus a practical and sufficiently accurate approximation for simulations of turbidity currents with fine sediments \citep{cantero2008history_model,xie2023auto,tsai2025history_model}. Inter-particle collisions are likewise neglected because the suspension is dilute and weakly inertial \citep{sundaram1997collision}.

The retained particle forces are coupled to the carrier flow through their reaction forces, replacing the solutal buoyancy term in \eqref{eq:solutal fluid momentum} as the forcing that drives the current. Under the dilute point-particle approximation, the governing equations of the carrier fluid are

\begin{align}
\label{eq:initial continuity}
  \nabla\cdot{\vec u}  &=0,\\
  \label{eq:initial fluid momentum}
{\partial_t \vec{u}}+{\vec u} \cdot \nabla {\vec u}&=-\nabla{p}^*+\nu\nabla^2\vec{u}+{\vec M}_{i},
\end{align}

where ${\vec M}_{i}$ is the momentum source term arising from particle feedback, modelled following \cite{climent2006coupling} as

\begin{equation}
   {\vec M}_{i}=\rho_f^{-1}\sum_{n=1}^{N_p}\left[m_p\left(\vec{g}-\frac{\d\vec{v}}{\d t}^{(n)}\right)-m_f\left(\vec{g}-\frac{\D\vec{u}}{\D t}(\vec x_p^{(n)})\right)\right] \mathcal{G}({\vec x}-{\vec x}_p^{(n)}),
\label{eq:Mi}
\end{equation}

where $N_p$ is the total number of particles and the superscript $(n)$ denotes the $n$th particle. The first term in the square bracket is the reaction to the surface force on a particle, while the second term accounts for the correction related to the point-particle approximation. This formulation therefore yields a feedback force comprising only $\vec{F}_d, \vec{F}_l$ and $\vec{F}_{ad}$. The kernel function $\mathcal{G}(\vec{x})$ is defined as
\begin{equation}
    \mathcal{G}(\vec{x}) = (2\pi\sigma^2)^{-3/2}\exp\left(-\frac{{\vec x}^2}{2\sigma^2}\right), 
\end{equation}
where $\sigma$ is the standard deviation and is set equal to the isotropic mesh spacing $\Delta x$. 
\subsection{Simulation Set-up}
\label{sec:setup}
The computational domain and boundary conditions follow those of the DNS of temporal inclined gravity currents reported by \cite{cui2025}. Periodic boundary conditions are imposed in the streamwise and spanwise directions, while the bottom boundary is no-slip and the top boundary free-slip. Particles reaching the bottom are removed from the active suspension and recorded as deposited. No subsequent resuspension is modelled, as the Shields number \citep{shields1936} remains relatively small such that significant resuspension is not expected in the low-grain-Reynolds-number regime ($Re_g=u_\tau d_p/\nu\lesssim1$) considered in this paper \citep{buffington1997erosion,paphitis2001erosion,loiseleux2005erosion,ouriemi2007erosion}. Here, $u_\tau=\left(\nu\left.\partial_z\overline u\right|_{z=0}\right)^{1/2}$ is the friction velocity. The dynamics of the deposited bed is beyond the scope of the present study.

In the precursor solutally-driven simulation, the heavier layer is initialised with a thickness $h_0$, buoyancy $b_0$ and velocity $u_0$, giving an initial bulk Richardson number $Ri_0 = -b_0h_0\cos\alpha/u_0^2 = 1.11$ and bulk Reynolds number $Re_0 = u_0h_0/\nu = 2500$. The simulation is carried out for a duration of  $140t^*$ to reach a fully developed state, where $t^*=h_0/\sqrt{-b_0h_0\sin\alpha}$ is a typical timescale.

In the inertial-particle simulations, the initial particle concentration field $c_0$ is prescribed such that the local buoyancy matches that of the precursor solutally-driven current $b_s$, namely 
\begin{equation}
    c_0 \left(1 - \rho_p/\rho_f\right) g = b_s,
\end{equation}
 In each case, we fix $\rho_p/\rho_f=3$ and  $b_s$ (and hence $c_0$) and vary only the particle diameter $d_p$; therefore the initial number of particles $N_{p0}$ increases with decreasing $d_p$. The characteristic volume concentration remains below $1\%$ in all cases. The ratio of $d_p$ to the Kolmogorov length scale $\eta_k=(\nu^3/\varepsilon_T)^{1/4}$ is kept below $0.5$. Here, $\varepsilon_T=h^{-1}\int \varepsilon \d z$ is the characteristic dissipation rate of TKE, where $h= \left(\int \overline{u} \d z\right)^2/\int \overline{u}^2 \d z$. The detailed simulation parameters are listed in table \ref{tab:sim}. The inertial-particle cases are labelled S31 to S08 in order of decreasing $d_p/\eta_k$. The two passive-tracer cases are S00 and S00U; each contains the same number of tracers as particles in S15 and is seeded either following the buoyancy field (S00) or uniformly in the domain (S00U). In all cases, the dispersed-phase velocity is initialised to match the local fluid velocity. The bulk flow Reynolds number at the time the Lagrangian phase is initialised is $Re_{ic}=\left.\int \overline{u}\d z\right/\nu=21600$.

The domain size is $(L_x,L_y,L_z)=(20h_0,10h_0,20h_0)$, discretised on a uniform $1024\times512\times1024$ grid. The vertical extent exceeds twice the current height, so the free-slip top boundary has a negligible influence on the flow. The periodic streamwise ($x$) and spanwise ($y$) extents correspond to approximately 6000 and 3000 wall units, respectively. The adequacy of the grid resolution and domain size is demonstrated in appendix~\ref{sec:resolution}. Further numerical details are provided by \citet{van2018smallscale,van2019,cui2025}. All simulations use the in-house DNS code SPARKLE \citep{craske2015jet1,nair2023sparkle}. Statistics are collected over an interval $t_{stat} \approx 15 \tau_k \approx t^*$, following an initial transient of approximately $30 \tau_k$ to allow particles to adjust to the turbulent flow. Although the currents evolve slowly over the sampling interval because of entrainment and the height-dependent particle flux, the reported profiles are quasi-stationary when expressed in the scaled vertical coordinate introduced below. Statistical convergence is further assessed by comparing profiles of mean particle velocity $\tilde{w}_p$ (with the averaging operator $\tilde{\cdot}$ defined in \eqref{eq:ave}) obtained over intervals of $10\tau_k$ and $15\tau_k$ for the largest-particle case S31. The mean absolute difference between the two profiles is $3.7\%$ of the mean magnitude of the longer-window profile, and the maximum local relative difference reaches $8.2\%$ near the dilute upper edge. These small differences support satisfactory statistical convergence.

\begin{table}
  \begin{center}
\def~{\hphantom{0}}
 \begin{tabular}{lcccccccccc}
Sim. 

& $d_p/\eta_k$
& $\rho_p/\rho_f$  
& $N_{p0}(10^9)$ 
& ${Re_{p\infty}}$ 
& ${St}(10^{-2})$ 
& ${St^+}(10^{-2})$ 
& $\beta$
& $\gamma$\\ [3pt]
S31   & 0.31  & 3 &0.3  & 0.394  & 1.8 &21.1  &0.40& 0.68 \\
S23   & 0.23   & 3& 0.8 & 0.172 & 1.0 &11.9  &0.23&1.16 \\
S15   & 0.15  & 3 & 2.7 & 0.052  &  0.5 &5.3 &0.11&2.55  \\
S08   & 0.08   & 3& 21.6  & 0.007   & 0.1 &1.2 &0.03&9.96  \\
S00   & -- &  --& 2.7  & 0 & 0  & 0 & 0&$\infty$ \\
S00U   & -- &  --& 2.7  & 0 & 0  & 0 & 0&$\infty$ \\
\end{tabular}
  \caption{Simulation details. Domain size for all simulations is $20h_0(x)\times 10h_0(y)\times20h_0(z)$ at a resolution of $1024\times 512\times1024$. The slope angle is $\alpha = \pi/180$ and the friction Reynolds number is $Re_\tau =530$ for all cases.}
  \label{tab:sim}
  \end{center}
\end{table}

\subsection{Theoretical framework}
\label{sec:theory}

The inner and outer layers of turbidity currents have distinct dynamics, and they are naturally separated by the level of $z_{um}$, where  $\overline{u}$ attains its maximum. Our primary interest is particle settling in the outer layer where the fluid velocities sampled at particle positions show pronounced positive values (see details further below), and hence we define the outer-layer characteristic streamwise velocity $u_{To}$,  concentration $c_{To}$ and turbulence kinetic energy (TKE) $e_{To}$ as \cite{cui2025}
\refstepcounter{equation}
\begin{equation*}
    u_{To} = h_o^{-1}\int_{z_{um}}^\infty \overline{u} \d z,\quad c_{To}=h_o^{-1}\int_{z_{um}}^\infty \overline{c} \d z,\quad e_{To}= h_o^{-1}\int_{z_{um}}^\infty e \d z,
   \tag{\theequation a-c}
\end{equation*}
where $h_o= (\int_{z_{um}}^\infty \overline{u} \d z)^2/(\int_{z_{um}}^\infty \overline{u}^2 \d z)$ is the characteristic thickness of the outer layer and $e=\overline{u_i'^2}/2$ is TKE. The averaging operator $\overline{(\cdot)}$ and the deviation from the average $(\cdot)'$ for an arbitrary quantity $\chi$ are defined as
\begin{equation}
    \overline{\chi}(z,t)= \frac{1}{L_x L_y} \iint \chi(\vec{x},t)\d x\d y,\qquad \chi' = \chi - \overline{\chi}.
    \label{eq:Reave}
\end{equation}
We define the terminal particle Reynolds number $Re_{p\infty}$, the bulk and wall-based Stokes numbers $St$ and $St^+$, and the friction Reynolds number $Re_\tau$ as
\refstepcounter{equation}\label{eq:basic_para}
\begin{equation*}
    Re_{p\infty} = \frac{d_p w_{p\infty}}{\nu},\qquad St =\frac{\tau_{st}}{\tau_k},\qquad St^+ = \frac{\tau_{st}u_\tau^2}{\nu}, \qquad Re_\tau = \frac{u_\tau z_{um}}{\nu},
   \tag{\theequation a-d}
\end{equation*}
where $w_{p\infty}$ is the terminal settling velocity in a quiescent fluid, defined as the steady-state solution of the particle force balance under buoyancy and nonlinear drag forces, $\tau_{st}= d_p^2(\rho_p/\rho_f)(18\nu)^{-1}$ is the Stokes relaxation time, $\tau_k=\sqrt{\nu/\varepsilon_{T}}$ is the Kolmogorov timescale.
We also define the settling-to-turbulence velocity ratio $\beta$ and the auto-suspension index $\gamma$ as
\refstepcounter{equation}\label{eq:settle_para}
\begin{equation*}
    \beta = \frac{{w_{p\infty}}}{\sqrt{e_{To}}},\qquad
    \gamma = \frac{u_{To}\alpha}{w_{p\infty}}.
   \tag{\theequation a-b}
\end{equation*}
These two parameters characterise outer-layer suspension in turbidity currents. A smaller $\beta$ indicates that turbulent velocity fluctuations are strong relative to settling, while $\gamma$ measures, in the small-slope limit, the ratio of the energy supplied to the current to that required to maintain sediment suspension \citep{bagnold1962auto,xie2023auto}. 

Table \ref{tab:sim} lists the parameters in \eqref{eq:basic_para} and \eqref{eq:settle_para} for each case. The small values of $St$ identify a weakly inertial fine-sediment regime relevant to long-runout turbidity currents, whereas $\beta$ and $\gamma$ show that settling remains dynamically relevant and auto-suspension may occur. The relatively larger values of $St^+$ for the coarser particles indicate that inertia becomes more important near the wall, allowing us to assess how it modifies the low-inertia baseline.

To connect the Lagrangian and Eulerian frameworks, we define a coarse-grained intrinsic average $\hat \cdot$ and its concentration-weighted average $\tilde \cdot$ for an arbitrary particle quantity $\phi$:

\refstepcounter{equation}\label{eq:ave}
\begin{equation*}
      \hat{\phi}(\vec x,t) = \frac{1}{c(\vec x,t)}\sum_{n=1}^{N_{p}} \phi^{(n)}V_p^{(n)}\mathcal{G}({\vec x}-{\vec x}_p^{(n)}),\qquad
  \tilde\phi(z,t) =\frac{\overline{c(\vec x,t)\hat{\phi}(\vec x,t)}}{\overline{c}(z,t)},
   \tag{\theequation a,b}
\end{equation*}
where the particle concentration $c(\vec x,t)$ is defined as $c= \sum_{n=1}^{N_{p}} V_p^{(n)}\mathcal{G}({\vec x}-{\vec x}_p^{(n)})$ and $V_p$ is the particle volume. 
The averaging operator $\tilde\cdot$ in (\ref{eq:ave}) is analogous to the Favre average \citep{favre1965ave} commonly used in density-varying flows.

For field variables $\phi$, we define $\phi^{(n)} \equiv \phi (\vec x_p^{(n)})$. Note that in this case, given the compact support of $\mathcal G$, we have 
\begin{equation}
  \overline{c}\tilde{\phi} =
  \overline{c \hat\phi} = \overline{\sum_{n=1}^{N_p}  \phi(\vec x_p^{(n)}) V_p^{(n)} \mathcal{G}(\vec x - \vec x_p^{(n)})}
  \approx \overline{\phi(\vec x) \sum_{n=1}^{N_p}   V_p^{(n)} \mathcal{G}(\vec x - \vec x_p^{(n)})}
  = \overline{c \phi},
  \label{eq:cphi}
\end{equation}
which becomes exact in the limit of $\sigma \rightarrow 0$.

\section{Dynamics of particle settling}\label{sec:dynamics}
\subsection{Decomposition of the effective settling velocity} \label{sec:decomp}

Despite the complexity of particle motion in turbulence, the particle velocity can be viewed as the local fluid velocity modified by a slip velocity between the particle and the surrounding flow. In the slope-normal direction this relation is
\begin{equation}
    w_p(\vec{x}_p,t) = w(\vec{x}_p,t)- w_s(\vec{x}_p,t),
    \label{eq:velrel}
\end{equation}
where $w_s(\vec{x}_p,t)$ is the fluid--particle slip velocity. Applying the averaging procedure defined in \eqref{eq:ave} yields
\begin{equation}
    \tilde w_p(z,t) = \tilde w(z,t) - \tilde w_s(z,t).
    \label{eq:velrel_ave}
\end{equation}
 Here, $\tilde w$ is the concentration-weighted mean fluid velocity sampled at particle positions. For the present flow, the Eulerian mean vertical fluid velocity $\overline{w}=0$ under the point-particle approximation, so any non-zero value of $\tilde w$ must arise from biased sampling of the turbulent field. Here, we use \emph{biased sampling} broadly to denote a difference between the mean fluid velocity sampled by the dispersed phase and the Eulerian mean, irrespective of the underlying mechanism. We reserve \emph{preferential sampling} for the selective sampling of particular turbulent flow structures by inertial particles. A negative value of $\tilde w$ implies that turbulence enhances settling through downward-directed effects, e.g., preferential sweeping as discussed in \S\ \ref{sec:Intro} or feedback-induced settling enhancement \citep{bosse2006s2way_enahnce,tom2022settling}, whereas a positive value indicates turbulence tends to suspend particles.
 
 The mean slip velocity $\tilde w_s$ in \eqref{eq:velrel_ave} quantifies the mean particle--fluid velocity difference. To identify the physical contributions to $\tilde w_s$, we rewrite the particle equation of motion in the slope-normal direction. Using the drag formulation in (\ref{eq:forces}$a$), \eqref{eq:dvdt} becomes
\begin{equation}
\frac{\d w_p}{\d t} = w_s\frac{\mathcal{F}(Re_p)}{\tau_{st}} + \frac{{F}_{l}^z+{F}_{ad}^z+{F}_{pv}^z}{m_p} - g^*\cos\alpha,
\label{eq:ws_ana}
\end{equation}
where $g^* = g(1-\rho_f/\rho_p)$ is the reduced gravity. 
Rearranging \eqref{eq:ws_ana} for $w_s$ and applying \eqref{eq:ave} yield three contributions to $\tilde{w}_s$:
\begin{equation}
\tilde{w}_s = w_{st}\widetilde{\frac{1}{\mathcal{F}(Re_p)}}+{\tau_{st}}
\widetilde{\frac{\dot{w}_p}{\mathcal{F}(Re_p)}}
-\tau_{st}\widetilde{\frac{f_{ap}}{\mathcal{F}(Re_p)}},
\label{eq:ws_ave}
\end{equation}
where $w_{st}=\tau_{st}g^*\cos\alpha$ is the Stokes settling velocity, $\dot{w}_p=\d w_p/\d t$ represents the departure from local force equilibrium, and $f_{ap}=(F_l^z+F_{ad}^z+F_{pv}^z)/m_p$ represents the acceleration due to additional hydrodynamic forces. The first two contributions have direct counterparts in the kinetic framework of \citet{bragg2021wallsweep}, derived from the first moment of the phase-space probability density function (PDF). In the Stokes-drag limit $\mathcal{F}(Re_p)=1$, they further resolved the force-nonequilibrium contribution into a mean-acceleration term associated with gradients in the mean particle velocity ($R_1$), concentration-gradient-driven inertial diffusion ($R_2$), and turbophoretic drift ($R_3$) in their equation (12).

\begin{figure}
  
  \centerline{\includegraphics[scale =1]{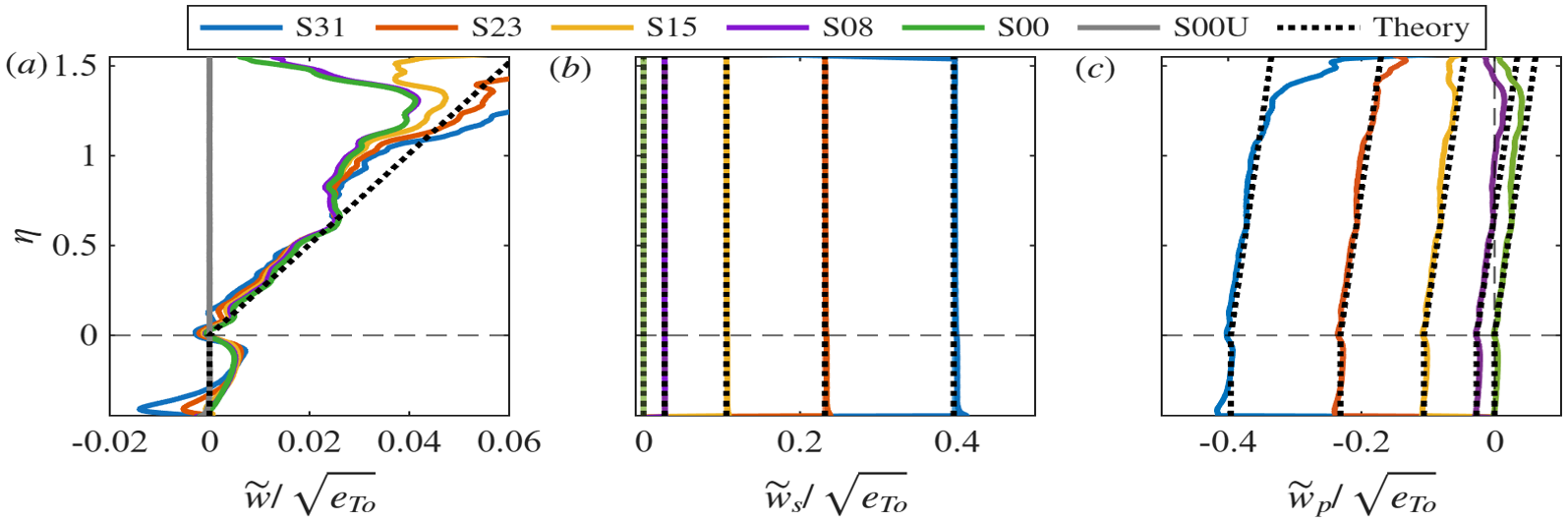}}
  \caption{Profiles of $(a)$ ${\tilde w}$, $(b)$  ${\tilde w_s}$, and $(c)$  ${\tilde w_p}$, together with $w_{p\infty}$ (dashed lines in panel $(b)$, shown in the same colour scheme as the labels but indistinguishable from $\tilde w_s$). The predictions from \eqref{eq:w_model}, \eqref{eq:ws_model} and \eqref{eq:finalModel} are shown by black dotted lines in panels $(a)$, $(b)$ and $(c)$, respectively. All quantities are normalised by $\sqrt{e_{To}}$ and plotted against $\eta$. The horizontal axis in panel $(c)$ is reversed relative to panel $(b)$.}
  \label{fig:velocities}
\end{figure}

Figure \ref{fig:velocities} shows the mean velocity components in \eqref{eq:velrel_ave}, normalised by $\sqrt{e_{To}}$, as a function of $\eta = (z-z_{um})/h_o$. The inner and outer layers are delineated by $\eta = 0 ~(z = z_{um})$. Panel $(a)$ shows the mean fluid velocity sampled at particle positions $\tilde{w}$. The passive-tracer case S00 exhibits $\tilde w>0$ throughout the depth, particularly in the outer layer, indicating an upward sampling bias even though the inertia-free tracers simply follow the zero-mean turbulent fluctuations. However, in the uniformly seeded case S00U with the same background flow field (solid grey line), this bias disappears and $\tilde w=0$ is preserved throughout. The results of these two passive-tracer cases suggest that the upward bias is directly linked to the concentration gradient, a point we revisit in the next section. For the inertial-particle cases, the profiles of $\tilde{w}$ in the outer layer nearly collapse onto the S00 profile, indicating that particle inertia has only a weak influence over the parameter range considered. In the inner layer, $\tilde{w}$ is generally small in magnitude for all cases, as examined in more detail below.

Figure \ref{fig:velocities} ($b,c$) shows the profiles of normalised ${\tilde w_s}$ and ${\tilde w_p}$ as functions of $\eta$. Panel ($b$) also includes the corresponding values of $w_{p\infty}$, shown as vertical dashed lines. The close coincidence of ${\tilde w_s}$ and $w_{p\infty}$ indicates that the leading-order force balance in the $z$-direction remains between buoyancy and drag. Accordingly, contributions from force-nonequilibrium and the additional hydrodynamic forces in \eqref{eq:ws_ave} are small in the weakly inertial regime considered here. The slight near-bed increase of $\tilde{w}_s$ (most evident for case S31 at $\tilde{w}_s/\sqrt{e_{To}}\approx0.4$ in figure \ref{fig:velocities} $(b)$) may be associated with turbophoretic drift, but its magnitude remains negligible compared with the settling contribution for the cases considered. For particles with $St>0.3$, however, turbophoresis may become the dominant settling-enhancement mechanism near the wall, as suggested by \cite{bragg2021wallsweep}. Meanwhile, the settling reduction due to nonlinear drag effects reported by \cite{good2014settling} is not observed here. The mean particle velocity $\tilde w_p$ shown in figure \ref{fig:velocities} ($c$) reflects the combined contributions of the biased sampling and the particle--fluid slip. For all inertial-particle cases, $\tilde w_p$ remains predominantly negative despite $\gamma>1$ for the smaller-particle cases, although it becomes progressively less so as $\gamma$ increases and the slip contribution diminishes with decreasing particle size.

\begin{figure}
  \centerline{\includegraphics[scale =1]{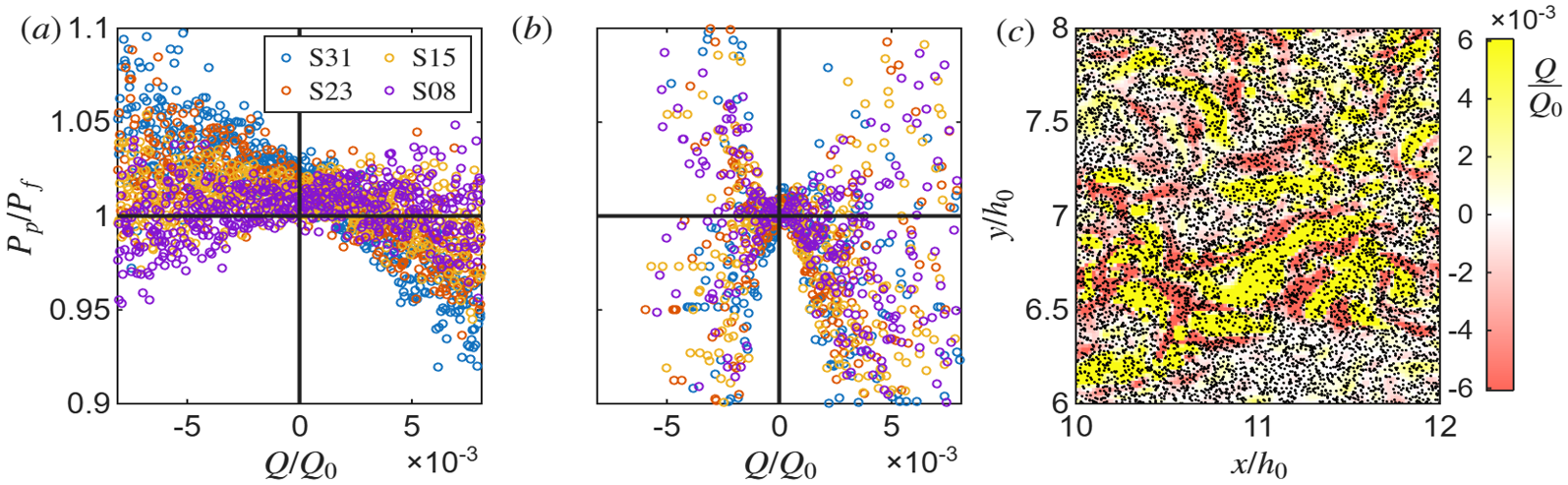}}

  \caption{Preferential-sampling ratio $P_p/P_f$ as a function of the normalised $Q$ criterion, $Q/Q_0$, at $(a)$ $z^+\approx60$ and $(b)$ $\eta\approx0.6$. The horizontal and vertical black lines denote $P_p/P_f=1$ and $Q=0$, respectively. $(c)$ Instantaneous field of $Q/Q_0$ over the region $x/h_0\in[10,12]$ and $y/h_0\in[6,8]$ at $z^+=60$ for case S31 at $t=40\tau_k$, with particle positions indicated by black points. }
  \label{fig:pdf}
\end{figure}

Markedly negative values of $\tilde{w}$ are found in the near-wall region for the larger-particle cases as shown in figure \ref{fig:velocities} ($a$). Although the bulk Stokes numbers $St$ for the cases are small, the larger values of wall-based $St^+$ (table \ref{tab:sim}) suggest a greater influence of particle inertia near the wall, where preferential sweeping may contribute to the negative values of $\tilde w$. To examine this possibility, we define the fluid- and particle-sampled PDFs of the $Q$ criterion, denoted by $P_f$ and $P_p$, respectively, as
\begin{equation}
P_f(q\mid z,t)=\overline{\delta\!\left[q-Q(\vec{x},t)\right]},\qquad
P_p(q\mid z,t)=\frac{\overline{c(\vec{x},t)\delta\!\left[q-Q(\vec{x},t)\right]}}{\overline{c}(z,t)},
\end{equation}
where $q$ denotes a realization of $Q$, and
\begin{equation}
Q=\frac{1}{8}\left(
\left|\nabla\vec{u}-\nabla\vec{u}^{T}\right|^2
-
\left|\nabla\vec{u}+\nabla\vec{u}^{T}\right|^2
\right).
\end{equation}
Here, $Q<0$ and $Q>0$ correspond to strain-dominated and rotation-dominated regions, respectively. The ratio $P_p/P_f$ therefore measures preferential sampling: $P_p/P_f=1$ denotes unbiased sampling, whereas values above unity indicate oversampling of a given range of $Q$. For inertial particles, centrifugal effects expel particles from vortical regions and are therefore associated with preferential sampling of strain-dominated regions.

Figures \ref{fig:pdf}$(a,b)$ compare $P_p/P_f$ at $z^+=zu_\tau/\nu\approx60$ in the inner layer and at $\eta\approx0.6$ in the outer layer, respectively, as functions of the normalised $Q$ criterion, $Q/Q_0$, where $Q_0=u_0^3/(h_0\nu)$. At $z^+\approx60$, the larger-particle cases preferentially sample $Q<0$, with $P_p/P_f>1$ over much of the strain-dominated range and falling below unity for sufficiently positive $Q$. This preference strengthens with increasing particle size, suggesting stronger centrifugal effects that may partly explain the negative near-wall $\tilde w$ observed for the larger particles. A qualitative indication of this tendency can also be seen in the instantaneous field for case S31 in figure \ref{fig:pdf}$(c)$, where particles within the displayed region show a preference for strain-dominated regions with $Q<0$. In contrast, at $\eta \approx 0.6$ in figure \ref{fig:pdf} $(b)$, $P_p/P_f$ fluctuates around unity without a systematic dependence on the sign of $Q$ for all cases, consistent with the weak role of particle inertia in the outer layer discussed above. 

Another mechanism that could potentially contribute to a negative $\tilde{w}$ in the inner layer is settling enhancement induced by particle feedback on the fluid. Because particle feedback provides the forcing that drives the turbidity current, a dynamically equivalent one-way-coupled reference case cannot be constructed. The independent effect of two-way coupling therefore cannot be determined from the present simulations. Nevertheless, the close agreement of cases S15 and S08 with the passive-tracer case (S00) as shown in figure \ref{fig:velocities}$(a)$ suggests that particle feedback has only a limited influence on $\tilde{w}$, at least for the two smallest particle sizes considered. 

\subsection{Modelling of the effective settling velocity}
\label{sec:derivate}
Armed with the decomposition in \eqref{eq:velrel_ave}, we now develop a theoretical model for $\tilde w_p$ by treating $\tilde w$ and $\tilde w_s$ separately. We first address the velocity component, $\tilde w$. Substituting $\phi=w$ into \eqref{eq:cphi}, we obtain

\begin{equation}
  \overline{c}\tilde w = {\overline{\hat w c}} \approx {\overline{ w c}} = {\overline{w}~\overline c} + {\overline{ w' c'}}.
  \label{eq:tilde_w}
\end{equation}
In the limit $\sigma \rightarrow 0$, \eqref{eq:tilde_w} can be written as 
\begin{equation}
    \overline{c}\tilde w = \overline{w}\,\overline c-K_c\partial_z\overline{c},
    \qquad
    K_c=-\overline{w'c'}\left(\partial_z\overline c\right)^{-1}.
    \label{eq:cw_kc}
\end{equation}
Here $K_c$ is defined from the measured turbulent particle flux. It therefore includes the effects of particle inertia on this flux, although we refer to it as a turbulent diffusivity for simplicity. This form is closely related to the more general kinetic framework summarised by \citet{bragg2021wallsweep}. Under the commonly used Gaussian approximation, their framework decomposes the sampled-fluid-velocity flux into drift and diffusive contributions
\begin{equation}
    \overline c\,\tilde w=\zeta\,\overline c-D^{[1]}\partial_z\overline c,
\end{equation}
where $\zeta$ is a concentration-gradient-independent drift coefficient and $D^{[1]}$ is the first-order diffusion coefficient. In their formulation, preferential sweeping is formally associated with the drift term. The present formulation \eqref{eq:cw_kc} instead follows directly from Reynolds and concentration-weighted averaging and does not assign the resulting terms to distinct physical mechanisms. This naturally links the particle-sampled fluid velocity to conventional Eulerian transport fluxes and enables its modelling using standard turbulence-transport arguments.

\begin{figure}
  \centerline{\includegraphics[scale =1]{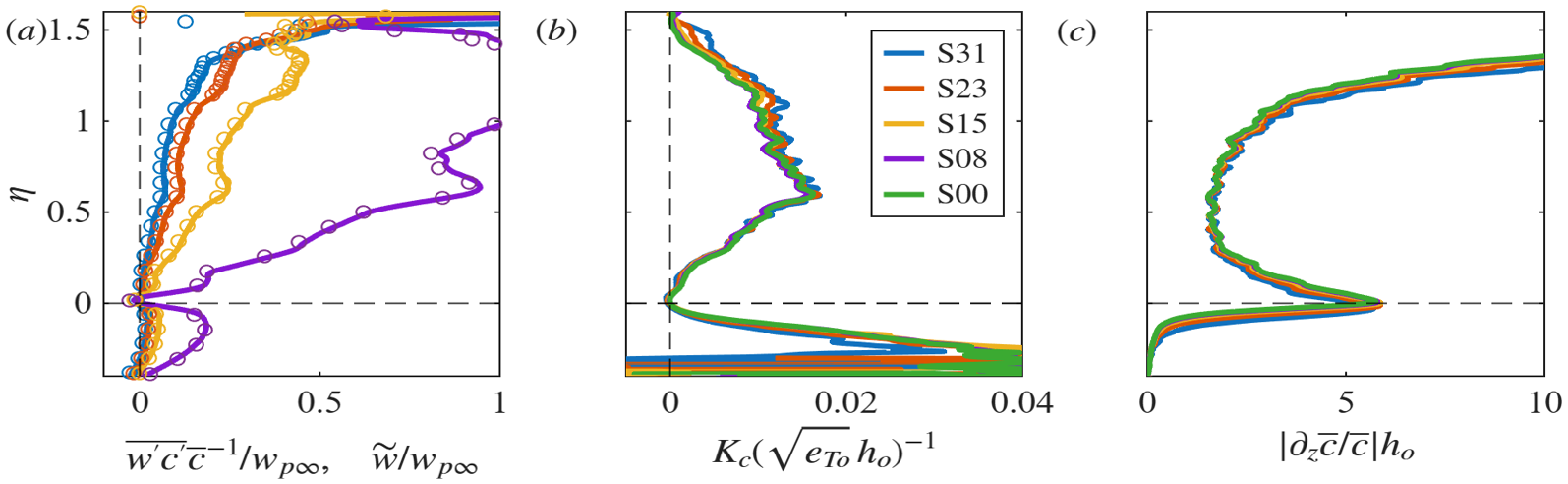}}
  \caption{Profiles of normalised $(a)~\overline{w'c'}/\overline{c}$ (solid lines), $\tilde w$ (circles), $(b)$ turbulent diffusivity $K_c$, and $(c)$ relative concentration gradient $|\partial_z\overline{c}/\overline{c}|$, plotted against $\eta$.}
  \label{fig:mixing}
\end{figure}

Substituting $\overline{w}=0$ into \eqref{eq:tilde_w}, we obtain
 \begin{equation}
    \tilde w \approx \frac{\overline{w'c'}}{\overline{c}}=-K_c \frac{\partial_z\overline{c}}{\overline{c}}.
    \label{eq:wkc}
\end{equation}
Figure \ref{fig:mixing} $(a)$ compares $\overline{w'c'}/\overline{c}$ (solid lines) with the directly sampled $\tilde w$ (circles), both normalised by $w_{p\infty}$, showing that the two quantities almost collapse for the inertial-particle cases. This confirms that the sampling bias is directly represented by the turbulent particle flux.  Figure \ref{fig:mixing} $(b)$ shows the normalised $K_c$. The close collapse of $K_c$ across the different cases suggests that particle inertia has little influence on the turbulent diffusivity over the parameter range considered, which is therefore governed primarily by the carrier flow dynamics. Near the shear-free velocity maximum ($\eta=0$), the vanishing mean shear weakens turbulence and the associated transport, leading to the small magnitude of $K_c$. Although the relative concentration gradient $|\partial_z\overline{c}/\overline{c}|$ shown in figure \ref{fig:mixing} $(c)$ is appreciable near the maximum, the magnitude of $\tilde w$ remains small. Farther below $\eta=0$, $K_c$ increases, but the concentration field is well mixed and the gradient becomes weak. The limited spatial overlap between these two quantities therefore restricts the magnitude of $\tilde w$ in the inner layer. Notably, the reduction of $K_c$ near $\eta=0$ may also act as a transport barrier between the inner and outer layers \citep{selfsharp2019, salinas2021nature}, which may support the maintenance of a dense basal driving layer \citep{luchi2018drivinglayer} and, in turn, the long-runout of turbidity currents \citep{wells2021ARFM}.

The dependence of $\tilde w$ on the concentration gradient and turbulent diffusivity indicates that the upward bias in the present cases arises from a simple gradient--diffusion process: more particles are transported into regions of lower concentration by upward turbulent motions than are returned by downward motions. This is also consistent with the uniformly seeded case S00U, which has no upward bias in the absence of a concentration gradient, despite having the same background turbulence as case S00. This further suggests, within the parameter space investigated here, that loitering-type effects \citep{nielsen1993loitering} are not required to produce the observed bias and are likely secondary to the gradient--diffusion mechanism. More broadly, this mechanism may also be relevant to other particle-laden flows. In wall-bounded turbulence, for example, near-wall particle accumulation produces a strong concentration gradient, which may contribute to the oversampling of ejection regions reported by \cite{marchioli2002Biased,bragg2021wallsweep,gao2024costalift}. From this perspective, such biased sampling need not arise solely from inertia-mediated interactions between particles and coherent turbulent structures, but may also result from turbulent transport acting on an inhomogeneous particle distribution. 

To model $\tilde{w}$ in the outer layer ($\eta>0$), we adopt the von Kármán--Pohlhausen-type integral framework used for fully developed gravity currents \citep{van2019,cui2025}. The mean velocity and concentration are approximated by linear profiles, while the TKE and Reynolds shear stress are represented by quadratic profiles:
\begin{equation}
    \overline{u}=u_{To}f_u(\eta),\quad
    \overline{c}=c_{To}f_c(\eta),\quad
    e=e_{To}f_e(\eta),\quad
    \overline{w'u'}=c_m e_{To}f_e(\eta),
\end{equation}
where $f_u=f_c=2(\eta_1-\eta)\eta_1^{-2}$, $f_e=6\eta(\eta_1-\eta)\eta_1^{-3}$, $\eta_1=4/3$, and $c_m\approx0.25$. Given the weak dependence of $K_c$ on particle inertia for the present cases, $K_c$ is modelled as
\begin{equation}
    K_c = \frac{K_m}{Sc_T}
    = -\frac{\overline{w'u'}}{Sc_T}\left(\frac{\partial \overline{u}}{\partial z}\right)^{-1}
    = \frac{8c_m e_{To}h_o}{9Sc_T u_{To}}f_e(\eta),
    \label{eq:Kc_model}
\end{equation}
where $K_m$ is the turbulent viscosity and $Sc_T\approx0.83$ is the turbulent Schmidt number. Substituting \eqref{eq:Kc_model} into \eqref{eq:wkc} gives
\begin{equation}
    \tilde w = c_w U_o \eta,\quad \eta>0,
    \label{eq:w_model}
\end{equation}
where $c_w=3c_m/(\eta_1 Sc_T)$ and $U_o=e_{To}/u_{To}$. Following equations (4.1$a$), (4.7) and (7.3) of \cite{cui2025}, together with equation (4.18) of \cite{van2019}, $U_o$ can be expressed as a function of the slope angle $\alpha$ and the depth-integrated buoyancy forcing $B=(\rho_p/\rho_f-1)g\sin\alpha\int \overline{c}\d z$:
\begin{equation}
    U_o(\alpha, B)
    = \frac{a_e}{a_u(\alpha)}
    \sqrt{
    \frac{B}
    {9(c_{um}-1)^2c_\kappa^2a_u^2(\alpha)/(4c_{um}^2)+1}
    },
    \quad
    a_u(\alpha)=
    \sqrt{
    \frac{8}{9}
    \frac{Sc_T+c_m/\tan\alpha}
    {Sc_T(c_m-c_\varepsilon)}
    },
\end{equation}
where $a_e\approx1.77$, $c_{um}\approx1.13$, $c_\varepsilon\approx0.21$ and $c_\kappa$ (von Kármán constant) $\approx0.41$ are constants.

Equation \eqref{eq:w_model} therefore predicts a linear dependence of $\tilde w$ on $\eta$ in the outer layer at $\alpha = 1^\circ$, as shown by the black dotted line in figure \ref{fig:velocities} $(a)$.  Although the closure for $K_c$ is derived from solutally-driven gravity current theory and does not account for the downward-directed effects (e.g., preferential sweeping), the model is able to capture the overall trends well, as these effects remain subdominant for the cases considered. However, the applicability of this closure to other flow conditions and particle properties requires further investigation.

The mean slip velocity $\tilde w_s$ closely matches the terminal velocity $w_{p\infty}$ obtained from the implicit force balance equation; however, we seek a simpler yet sufficiently accurate expression in terms of the \emph{a priori} particle Reynolds number $Re_{pst}=w_{st}d_{p}/\nu$, based on the Stokes settling velocity $w_{st}$. Neglecting the subdominant mean contributions in \eqref{eq:ws_ave}, and approximating $\widetilde{\frac{1}{\mathcal{F}(Re_p)}}$ by $\frac{1}{\mathcal{F}(Re_{pst})}$, we obtain
\begin{equation}
    \tilde w_s \approx \frac{w_{st}}{\mathcal{F}(Re_{pst})},
    \label{eq:ws_model}
\end{equation}
where $\mathcal{F}(Re_{pst})=1+0.15Re_{pst}^{0.687}$. Figure \ref{fig:velocities} $(b)$ shows that $w_{st}/\mathcal{F}(Re_{pst})$ (black dotted lines) collapses well with $\tilde w_s$, confirming the accuracy of this approximation. Combining \eqref{eq:w_model} and \eqref{eq:ws_model}, and noting that $\tilde w$ has relatively small magnitude in the inner layer (i.e.\ $\tilde w\approx 0$), we predict
\begin{equation}
  \tilde w_p =\tilde w - \tilde w_s= \left\{
    \begin{array}{ll}
      -w_{st}/\mathcal{F}(Re_{pst}), & \eta\le 0 \\[2pt]
      c_wU_o(\alpha,B)\eta-w_{st}/\mathcal{F}(Re_{pst}) ,         & \eta>0.
    \end{array} \right.
    \label{eq:finalModel}
\end{equation}
 Figure \ref{fig:velocities} $(c)$ compares this prediction (black dotted lines) with the simulation results and shows good agreement, although the present model does not incorporate inertia-related mechanisms that may enhance settling. For the cases considered here, the settling process can therefore be reasonably approximated as a competition between an effective velocity characterising down-gradient transport by turbulence and a settling velocity corrected for finite $Re_p$. This characterisation bears similarities to the classical Rouse framework \citep{rouse1939analysis}, although it is expressed here in terms of velocities rather than fluxes. However, the Rouse solution models the turbulent diffusivity using wall quantities, whereas the present closure is based on local flow properties that are more appropriate for capturing settling within the outer layer of turbidity currents. Notably, the predictive model is restricted to dilute, low-$St$, low-$Re_p$ conditions. Its accuracy may deteriorate when inertial effects, history effects or particle collisions become dynamically significant.

\section{Conclusion}\label{sec:conclusion}
In this study, we used Eulerian--Lagrangian DNS of solutally-driven currents seeded with passive tracers and inertial-particle-driven currents to examine the settling dynamics of dilute, mono-disperse particles in the inner and outer layers of turbidity currents. We decompose the particle velocity $w_p$ into the local fluid velocity $w$ and the particle--fluid slip velocity $w_s$, and map these Lagrangian quantities onto Eulerian profiles through concentration-weighted averaging of the coarse-grained fields. This framework directly links biased sampling of the turbulent velocity field to particle fluxes and identifies an upward bias arising from turbulent transport acting on an inhomogeneous particle distribution, without requiring particle inertia. For the weakly inertial regime considered here, downward-directed transport mechanisms such as preferential sweeping may modify this baseline behaviour, but appear to remain of secondary overall importance.

We show that although the vertical fluid velocity averaged over any horizontal plane ($\overline{w}$) is zero, the mean fluid velocity sampled at particle positions $\tilde w$ shows a pronounced upward bias in the outer layer. The DNS data and theoretical analysis show that this biased mean accords with the turbulent transport expected in an inhomogeneous concentration field, i.e.\ $\tilde w \approx \overline{w'c'}/\overline{c}$, and is well described by a gradient--diffusion model using the turbulent-diffusivity closure for solutally-driven gravity currents of \cite{cui2025}. Importantly, in the passive-tracer simulations, this biased mean is maintained only when a concentration gradient is present. We therefore infer that loitering \citep{nielsen1993loitering} or inertial effects are not required to produce the upward bias for the cases considered here.

In the inner layer, $\tilde w$ is generally weak because the turbulent diffusivity and concentration gradient have limited spatial overlap. Near the shear-free velocity maximum, the vanishing mean shear weakens turbulent transport, whereas stronger turbulence maintains a well-mixed concentration field with only a weak gradient throughout the bulk of the layer. In the near-wall region, the larger particles exhibit negative $\tilde w$, consistent with an increasing influence of inertia-related preferential sampling.

The mean slip velocity of particles $\tilde w_s$ can be decomposed into a settling contribution, a force-nonequilibrium contribution and contributions from additional hydrodynamic forces, consistent with the kinetic framework of \citet{bragg2021wallsweep}. For the present weakly inertial cases, however, $\tilde w_s$ satisfies a leading-order balance between drag and buoyancy, while particle acceleration and other forces, such as shear-induced lift and added mass, make only minor contributions to the overall dynamics. Therefore, $\tilde w_s$ is well approximated by the quiescent terminal settling velocity corrected for finite particle Reynolds number.

Upon combining our dynamical representation for $\tilde w$ and $\tilde w_s$, we develop an Eulerian predictive model for the effective settling velocity $\tilde w_p$ in a turbidity current. The model predicts that $\tilde w_p$ increases linearly with dimensionless height $\eta$ in the outer layer and is dominated by the slip contribution $-\tilde w_s$ in the inner layer. This formulation yields predictions that are in good agreement with the DNS data across all cases considered.

A limitation of the model is that it does not account for the downward-directed particle transport mechanisms, which remain subdominant for the range of cases considered here. Its accuracy may therefore deteriorate when these mechanisms become dynamically significant. Extending the modelling framework to incorporate these effects, particularly at higher inertia or mass loading, is a natural direction for future work. Experimental measurements that quantify the velocity components in \eqref{eq:velrel_ave}, the turbulent particle flux and the concentration gradient would provide valuable ground truth for testing the proposed framework, but such data are currently lacking. The significance of the transport barrier near the velocity maximum for the long-runout behaviour of turbidity currents \citep{wells2021ARFM} also warrants further investigation. More broadly, the model highlights the role of particle concentration in modulating the settling process, a mechanism that is likely to be relevant across a wider class of particle-laden flows. For instance, the oversampling of ejection regions in wall turbulence may, at least in part, be influenced by the background concentration gradient established by near-wall particle accumulation \citep{marchioli2002Biased}, rather than solely by inertia-mediated preferential sampling of turbulent structures.

\paragraph*{Funding}
{The authors acknowledge the UK Turbulence Consortium (EPSRC grant EP/R029326/1 and EP/X035484/1), for the grand challenge project that provided the computational resources for this work.}

\paragraph*{Declaration of interests}
{The authors report no conflict of interest.}

\paragraph*{Data availability statement}
{The data that support the findings will be made openly available upon publication.}

\clearpage
\appendix

\section{Assessment of grid resolution and domain size}
\label{sec:resolution}
\begin{figure}[htbp]
    \centering
    \centerline{\includegraphics[scale =0.1]{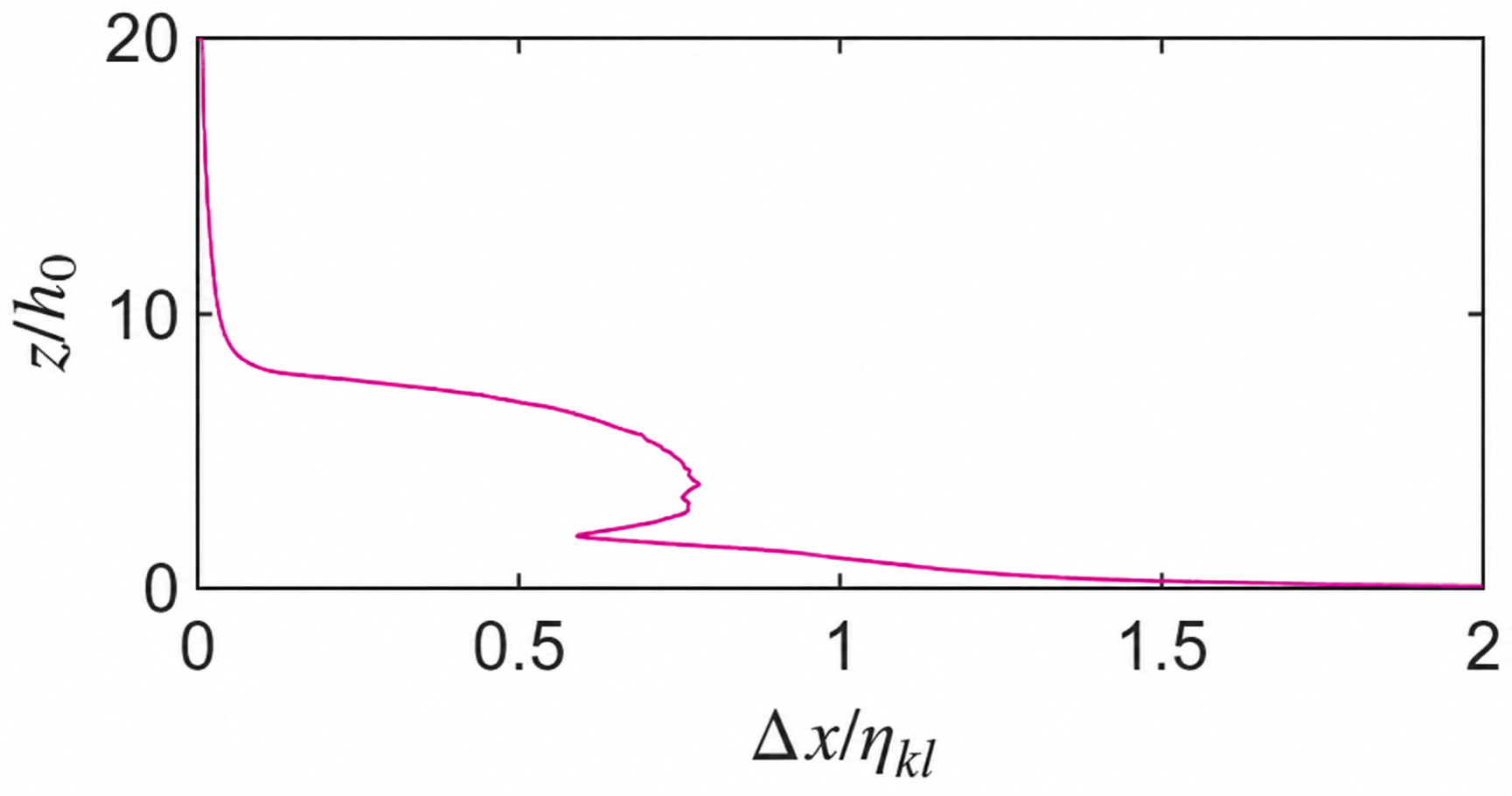}}
    \caption{Normalised grid spacing $\Delta x/\eta_{kl}$ at $t=30\tau_k$ for case S23.}
    \label{fig:app_kolmogorov}
\end{figure}

\begin{figure}[htbp]
    \centering
    \centerline{\includegraphics[scale =1]{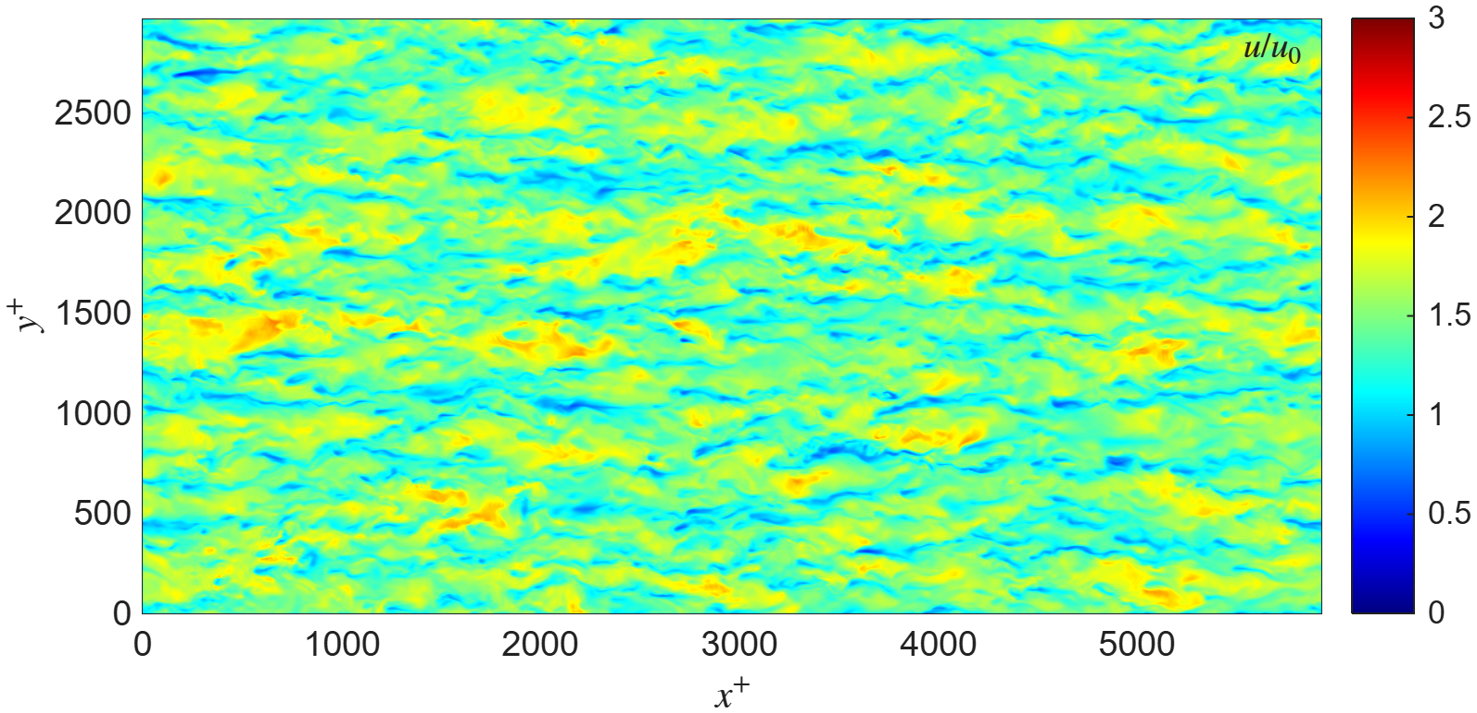}}
    \caption{Instantaneous streamwise velocity $u/u_0$
    at $z^+=30$ and $t=30\tau_k$ for case S23.}
    \label{fig:app_ufield}
\end{figure}

\begin{figure}[htbp]
    \centering
   \centerline{\includegraphics[scale =1]{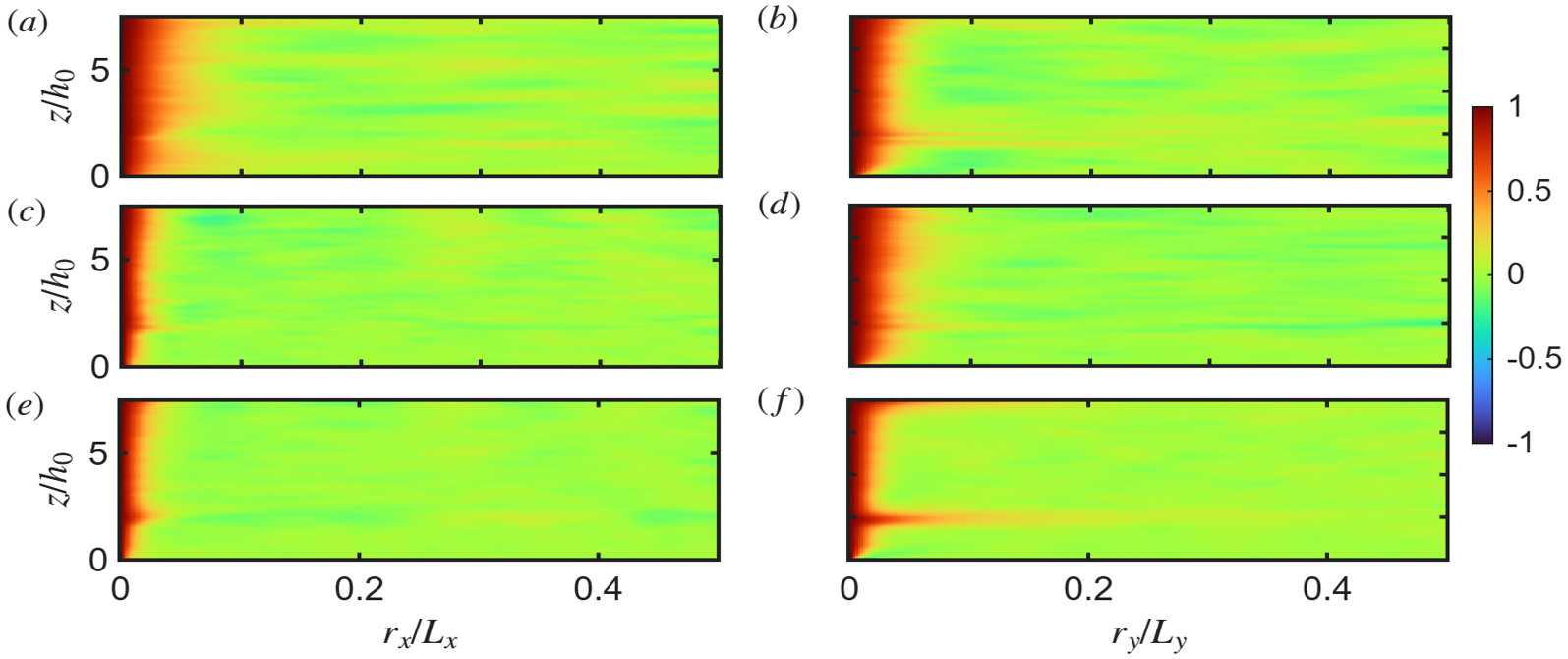}}
    \caption{Two-point velocity correlations
    $(a)$ $R_{uu}^{x}$, $(b)$ $R_{uu}^{y}$,
    $(c)$ $R_{vv}^{x}$, $(d)$ $R_{vv}^{y}$,
    $(e)$ $R_{ww}^{x}$ and $(f)$ $R_{ww}^{y}$
    for case S23 at $t=30\tau_k$.}
    \label{fig:app_corr}
\end{figure}

\suppressfloats[t]
The grid resolution is assessed using the local Kolmogorov length scale $\eta_{kl}=(\nu^3/\overline{\varepsilon})^{1/4}$. Figure~\ref{fig:app_kolmogorov} shows that the grid spacing satisfies $\Delta x/\eta_{kl}\leq2$ and remains below unity throughout the outer shear layer central to this study. Figure~\ref{fig:app_ufield} shows the instantaneous streamwise-velocity field at $z^+=30$ for case S23, revealing numerous distinct high- and low-speed streaks that are not constrained by the domain size in either periodic direction. To further assess the adequacy of the domain size, we plot the normalised two-point velocity correlations in figure~\ref{fig:app_corr}, which are defined as
\begin{equation}
R_{qq}^{\varrho}(r_\varrho,z,t)=
\frac{\overline{q'(\boldsymbol{x},t)
q'(\boldsymbol{x}+r_\varrho\boldsymbol{e}_\varrho,t)}}
{\overline{q'^2(\boldsymbol{x},t)}},
\qquad q\in\{u,v,w\},\quad \varrho\in\{x,y\}.
\end{equation}
The displaced point is evaluated with periodic wrap-around, and only the non-redundant interval $0\leq r_\varrho\leq L_\varrho/2$ is considered. All correlations decay to almost zero well before half of the corresponding domain length, indicating that the energy-containing
turbulent structures are adequately accommodated.

\clearpage
\bibliographystyle{plainnat}
\bibliography{references}

\begin{thebibliography}{48}
\providecommand{\natexlab}[1]{#1}
\providecommand{\url}[1]{\texttt{#1}}
\expandafter\ifx\csname urlstyle\endcsname\relax
  \providecommand{\doi}[1]{doi: #1}\else
  \providecommand{\doi}{doi: \begingroup \urlstyle{rm}\Url}\fi

\bibitem[Aliseda et~al.(2002)Aliseda, Cartellier, Hainaux, and
  Lasheras]{aliseda2002settling}
A.~Aliseda, A.~Cartellier, F.~Hainaux, and J.C. Lasheras.
\newblock Effect of preferential concentration on the settling velocity of
  heavy particles in homogeneous isotropic turbulence.
\newblock \emph{J. Fluid Mech.}, 468:\penalty0 77--105, 2002.

\bibitem[Bagnold(1962)]{bagnold1962auto}
R.A. Bagnold.
\newblock Auto-suspension of transported sediment; turbidity currents.
\newblock \emph{Proc. R. Soc. Lond. A: Math. Phys. Sci.}, 265\penalty0
  (1322):\penalty0 315--319, 1962.

\bibitem[Bec et~al.(2014)Bec, Homann, and Ray]{bec2014settling}
J.~Bec, H.~Homann, and S.S. Ray.
\newblock Gravity-driven enhancement of heavy particle clustering in turbulent
  flow.
\newblock \emph{Phys. Rev. Lett.}, 112\penalty0 (18):\penalty0 184501, 2014.

\bibitem[Bosse et~al.(2006)Bosse, Kleiser, and Meiburg]{bosse2006s2way_enahnce}
T.~Bosse, L.~Kleiser, and E.~Meiburg.
\newblock Small particles in homogeneous turbulence: settling velocity
  enhancement by two-way coupling.
\newblock \emph{Phys. Fluids}, 18\penalty0 (2):\penalty0 027102, 2006.

\bibitem[Bragg et~al.(2021{\natexlab{a}})Bragg, Richter, and
  Wang]{bragg2021wallsweep}
A.D. Bragg, D.H. Richter, and G.~Wang.
\newblock Mechanisms governing the settling velocities and spatial
  distributions of inertial particles in wall-bounded turbulence.
\newblock \emph{Phys. Rev. Fluids}, 6\penalty0 (6):\penalty0 064302,
  2021{\natexlab{a}}.

\bibitem[Bragg et~al.(2021{\natexlab{b}})Bragg, Richter, and
  Wang]{bragg2021wallsweep2}
A.D. Bragg, D.H. Richter, and G.~Wang.
\newblock Settling strongly modifies particle concentrations in wall-bounded
  turbulent flows even when the settling parameter is asymptotically small.
\newblock \emph{Phys. Rev. Fluids}, 6\penalty0 (12):\penalty0 124301,
  2021{\natexlab{b}}.

\bibitem[Brandt and Coletti(2022)]{brandt2022ARFM}
L.~Brandt and F.~Coletti.
\newblock Particle-laden turbulence: progress and perspectives.
\newblock \emph{Annu. Rev. Fluid Mech.}, 54\penalty0 (1):\penalty0 159--189,
  2022.

\bibitem[Buffington and Montgomery(1997)]{buffington1997erosion}
J.M. Buffington and D.R. Montgomery.
\newblock A systematic analysis of eight decades of incipient motion studies,
  with special reference to gravel-bedded rivers.
\newblock \emph{Water Resour. Res.}, 33\penalty0 (8):\penalty0 1993--2029,
  1997.

\bibitem[Cantero et~al.(2008)Cantero, Balachandar, and
  Garcia]{cantero2008history_model}
M.I. Cantero, S.~Balachandar, and M.H. Garcia.
\newblock An {Eulerian}--{Eulerian} model for gravity currents driven by
  inertial particles.
\newblock \emph{Intl J. Multiphase Flow}, 34\penalty0 (5):\penalty0 484--501,
  2008.

\bibitem[Caporaloni et~al.(1975)Caporaloni, Tampieri, Trombetti, and
  Vittori]{caporaloni1975turbo}
M.~Caporaloni, F.~Tampieri, F.~Trombetti, and O.~Vittori.
\newblock Transfer of particles in nonisotropic air turbulence.
\newblock \emph{J. Atmos. Sci.}, 32\penalty0 (3):\penalty0 565--568, 1975.

\bibitem[Climent and Magnaudet(2006)]{climent2006coupling}
E.~Climent and J.~Magnaudet.
\newblock Dynamics of a two-dimensional upflowing mixing layer seeded with
  bubbles: bubble dispersion and effect of two-way coupling.
\newblock \emph{Phys. Fluids}, 18\penalty0 (10):\penalty0 103304, 2006.

\bibitem[Craske and van Reeuwijk(2015)]{craske2015jet1}
J.~Craske and M.~van Reeuwijk.
\newblock Energy dispersion in turbulent jets. {Part} 1. direct simulation of
  steady and unsteady jets.
\newblock \emph{J. Fluid Mech.}, 763:\penalty0 500--537, 2015.

\bibitem[Cui et~al.(2025)Cui, Hughes, and van Reeuwijk]{cui2025}
L.~Cui, G.O. Hughes, and M.~van Reeuwijk.
\newblock Structure and scaling of inclined temporal gravity currents.
\newblock \emph{J. Fluid Mech.}, 1022:\penalty0 A34, 2025.

\bibitem[Daitche(2015)]{daitche2015history_dp_eta}
A.~Daitche.
\newblock On the role of the history force for inertial particles in
  turbulence.
\newblock \emph{J. Fluid Mech.}, 782:\penalty0 567--593, 2015.

\bibitem[Daitche and T{\'e}l(2011)]{daitche2011memory}
A.~Daitche and T.~T{\'e}l.
\newblock Memory effects are relevant for chaotic advection of inertial
  particles.
\newblock \emph{Phys. Rev. Lett.}, 107\penalty0 (24):\penalty0 244501, 2011.

\bibitem[Daitche and T{\'e}l(2014)]{daitche2014memory}
A.~Daitche and T.~T{\'e}l.
\newblock Memory effects in chaotic advection of inertial particles.
\newblock \emph{New J. Phys.}, 16\penalty0 (7):\penalty0 073008, 2014.

\bibitem[Dorrell et~al.(2019)Dorrell, Peakall, Darby, Parsons, Johnson, Sumner,
  Wynn, {\"O}zsoy, and Tezcan]{selfsharp2019}
R.M. Dorrell, J.~Peakall, S.E. Darby, D.R. Parsons, J.~Johnson, E.J. Sumner,
  R.B. Wynn, E.~{\"O}zsoy, and D.~Tezcan.
\newblock Self-sharpening induces jet-like structure in seafloor gravity
  currents.
\newblock \emph{Nat. Commun.}, 10\penalty0 (1):\penalty0 1381, 2019.

\bibitem[Favre(1965)]{favre1965ave}
A.J. Favre.
\newblock The equations of compressible turbulent gases.
\newblock Technical Report AD0622097, Defense Technical Information Center,
  1965.

\bibitem[Gao et~al.(2024)Gao, Shi, Parsani, and Costa]{gao2024costalift}
W.~Gao, P.~Shi, M.~Parsani, and P.~Costa.
\newblock On the relevance of lift force modelling in turbulent wall flows with
  small inertial particles.
\newblock \emph{J. Fluid Mech.}, 988:\penalty0 A47, 2024.

\bibitem[God{\'e} et~al.(2023)God{\'e}, Charton, Climent, and
  Legendre]{gode2023history_dp_eta}
H.~God{\'e}, S.~Charton, E.~Climent, and D.~Legendre.
\newblock {Basset}-{Boussinesq} history force acting on a drop in an
  oscillatory flow.
\newblock \emph{Phys. Rev. Fluids}, 8\penalty0 (7):\penalty0 073605, 2023.

\bibitem[Good et~al.(2014)Good, Ireland, Bewley, Bodenschatz, Collins, and
  Warhaft]{good2014settling}
G.H. Good, P.J. Ireland, G.P. Bewley, E.~Bodenschatz, L.R. Collins, and
  Z.~Warhaft.
\newblock Settling regimes of inertial particles in isotropic turbulence.
\newblock \emph{J. Fluid Mech.}, 759:\penalty0 R3, 2014.

\bibitem[Ling et~al.(2013)Ling, Parmar, and
  Balachandar]{ling2013history_dp_eta}
Y.~Ling, M.~Parmar, and S.~Balachandar.
\newblock A scaling analysis of added-mass and history forces and their
  coupling in dispersed multiphase flows.
\newblock \emph{Intl J. Multiphase Flow}, 57:\penalty0 102--114, 2013.

\bibitem[Loiseleux et~al.(2005)Loiseleux, Gondret, Rabaud, and
  Doppler]{loiseleux2005erosion}
T.~Loiseleux, P.~Gondret, M.~Rabaud, and D.~Doppler.
\newblock Onset of erosion and avalanche for an inclined granular bed sheared
  by a continuous laminar flow.
\newblock \emph{Phys. Fluids}, 17\penalty0 (10):\penalty0 103304, 2005.

\bibitem[Luchi et~al.(2018)Luchi, Balachandar, Seminara, and
  Parker]{luchi2018drivinglayer}
R.~Luchi, S.~Balachandar, G.~Seminara, and G.~Parker.
\newblock Turbidity currents with equilibrium basal driving layers: a mechanism
  for long runout.
\newblock \emph{Geophys. Res. Lett.}, 45\penalty0 (3):\penalty0 1518--1526,
  2018.

\bibitem[Magnaudet and Eames(2000)]{magnaudet2000forces}
J.~Magnaudet and I.~Eames.
\newblock The motion of high-{Reynolds}-number bubbles in inhomogeneous flows.
\newblock \emph{Annu. Rev. Fluid Mech.}, 32\penalty0 (1):\penalty0 659--708,
  2000.

\bibitem[Marchioli and Soldati(2002)]{marchioli2002Biased}
C.~Marchioli and A.~Soldati.
\newblock Mechanisms for particle transfer and segregation in a turbulent
  boundary layer.
\newblock \emph{J. Fluid Mech.}, 468:\penalty0 283--315, 2002.

\bibitem[Maxey(1987)]{maxey1987orginal}
M.R. Maxey.
\newblock The gravitational settling of aerosol particles in homogeneous
  turbulence and random flow fields.
\newblock \emph{J. Fluid Mech.}, 174:\penalty0 441--465, 1987.

\bibitem[Maxey and Riley(1983)]{maxey1983equation}
M.R. Maxey and J.J. Riley.
\newblock Equation of motion for a small rigid sphere in a nonuniform flow.
\newblock \emph{Phys. Fluids}, 26\penalty0 (4):\penalty0 883--889, 1983.

\bibitem[Nair et~al.(2023)Nair, Devenish, and van Reeuwijk]{nair2023sparkle}
V.~Nair, B.~Devenish, and M.~van Reeuwijk.
\newblock Effect of gravity on particle clustering and collisions in decaying
  turbulence.
\newblock \emph{Flow Turbul. Combust.}, 110\penalty0 (4):\penalty0 889--915,
  2023.

\bibitem[Nielsen(1993)]{nielsen1993loitering}
P.~Nielsen.
\newblock Turbulence effects on the settling of suspended particles.
\newblock \emph{J. Sediment. Res.}, 63\penalty0 (5):\penalty0 835--838, 1993.

\bibitem[Ouriemi et~al.(2007)Ouriemi, Aussillous, Medale, Peysson, and
  Guazzelli]{ouriemi2007erosion}
M.~Ouriemi, P.~Aussillous, M.~Medale, Y.~Peysson, and E.~Guazzelli.
\newblock Determination of the critical {Shields} number for particle erosion
  in laminar flow.
\newblock \emph{Phys. Fluids}, 19\penalty0 (6):\penalty0 061706, 2007.

\bibitem[Paphitis(2001)]{paphitis2001erosion}
D.~Paphitis.
\newblock Sediment movement under unidirectional flows: an assessment of
  empirical threshold curves.
\newblock \emph{Coast. Engng}, 43\penalty0 (3--4):\penalty0 227--245, 2001.

\bibitem[Reeks(1983)]{reeks1983turbo}
M.W. Reeks.
\newblock The transport of discrete particles in inhomogeneous turbulence.
\newblock \emph{J. Aerosol Sci.}, 14\penalty0 (6):\penalty0 729--739, 1983.

\bibitem[Rosa et~al.(2016)Rosa, Parishani, Ayala, and Wang]{rosa2016settling}
B.~Rosa, H.~Parishani, O.~Ayala, and L.-P. Wang.
\newblock Settling velocity of small inertial particles in homogeneous
  isotropic turbulence from high-resolution {DNS}.
\newblock \emph{Intl J. Multiphase Flow}, 83:\penalty0 217--231, 2016.

\bibitem[Rouse(1939)]{rouse1939analysis}
H.~Rouse.
\newblock \emph{An Analysis of Sediment Transportation in the Light of Fluid
  Turbulence}, volume~25.
\newblock US Department of Agriculture, Soil Conservation Service, Washington,
  DC, 1939.

\bibitem[Saffman(1965)]{saffman1965lift}
P.G. Saffman.
\newblock The lift on a small sphere in a slow shear flow.
\newblock \emph{J. Fluid Mech.}, 22\penalty0 (2):\penalty0 385--400, 1965.

\bibitem[Salinas et~al.(2021)Salinas, Balachandar, Shringarpure, Fedele, Hoyal,
  Zu{\~n}iga, and Cantero]{salinas2021nature}
J.S. Salinas, S.~Balachandar, M.~Shringarpure, J.~Fedele, D.~Hoyal,
  S.~Zu{\~n}iga, and M.I. Cantero.
\newblock Anatomy of subcritical submarine flows with a lutocline and an
  intermediate destruction layer.
\newblock \emph{Nat. Commun.}, 12\penalty0 (1):\penalty0 1649, 2021.

\bibitem[Schiller(1933)]{schiller1933Recor}
L.~Schiller.
\newblock {\"U}ber die grundlegenden {Berechnungen} bei der
  {Schwerkraftaufbereitung}.
\newblock \emph{Z. Ver. Dtsch. Ing.}, 77:\penalty0 318--321, 1933.

\bibitem[Shields(1936)]{shields1936}
A.~Shields.
\newblock Application of similarity principles and turbulence research to
  bed-load movement.
\newblock Technical report, California Institute of Technology, 1936.

\bibitem[Sundaram and Collins(1997)]{sundaram1997collision}
S.~Sundaram and L.R. Collins.
\newblock Collision statistics in an isotropic particle-laden turbulent
  suspension. {Part} 1. direct numerical simulations.
\newblock \emph{J. Fluid Mech.}, 335:\penalty0 75--109, 1997.

\bibitem[Tom and Bragg(2019)]{tom2019settling}
J.~Tom and A.D. Bragg.
\newblock Multiscale preferential sweeping of particles settling in turbulence.
\newblock \emph{J. Fluid Mech.}, 871:\penalty0 244--270, 2019.

\bibitem[Tom et~al.(2022)Tom, Carbone, and Bragg]{tom2022settling}
J.~Tom, M.~Carbone, and A.D. Bragg.
\newblock How does two-way coupling modify particle settling and the role of
  multiscale preferential sweeping?
\newblock \emph{J. Fluid Mech.}, 947:\penalty0 A7, 2022.

\bibitem[Tsai and Chou(2025)]{tsai2025history_model}
Y.-H. Tsai and Y.-J. Chou.
\newblock On the suspension and deposition within turbidity currents.
\newblock \emph{J. Fluid Mech.}, 1003:\penalty0 A1, 2025.

\bibitem[van Reeuwijk et~al.(2018)van Reeuwijk, Krug, and
  Holzner]{van2018smallscale}
M.~van Reeuwijk, D.~Krug, and M.~Holzner.
\newblock Small-scale entrainment in inclined gravity currents.
\newblock \emph{Environ. Fluid Mech.}, 18\penalty0 (1):\penalty0 225--239,
  2018.

\bibitem[van Reeuwijk et~al.(2019)van Reeuwijk, Holzner, and
  Caulfield]{van2019}
M.~van Reeuwijk, M.~Holzner, and C.P. Caulfield.
\newblock Mixing and entrainment are suppressed in inclined gravity currents.
\newblock \emph{J. Fluid Mech.}, 873:\penalty0 786--815, 2019.

\bibitem[Wang and Maxey(1993)]{wang1993settling}
L.-P. Wang and M.R. Maxey.
\newblock Settling velocity and concentration distribution of heavy particles
  in homogeneous isotropic turbulence.
\newblock \emph{J. Fluid Mech.}, 256:\penalty0 27--68, 1993.

\bibitem[Wells and Dorrell(2021)]{wells2021ARFM}
M.G. Wells and R.M. Dorrell.
\newblock Turbulence processes within turbidity currents.
\newblock \emph{Annu. Rev. Fluid Mech.}, 53\penalty0 (1):\penalty0 59--83,
  2021.

\bibitem[Xie et~al.(2023)Xie, Hu, Zhu, Yu, and P{\"a}htz]{xie2023auto}
J.~Xie, P.~Hu, C.~Zhu, Z.~Yu, and T.~P{\"a}htz.
\newblock Turbidity currents propagating down an inclined slope: particle
  auto-suspension.
\newblock \emph{J. Fluid Mech.}, 954:\penalty0 A44, 2023.

\end{thebibliography}

\end{document}